\documentclass[runningheads]{template/llncs}
\usepackage{multirow}
\usepackage{makecell}
\usepackage{graphicx}
\usepackage{comment}
\usepackage{url}
\usepackage[perpage]{footmisc}
\usepackage{xcolor}
\usepackage{hyperref}
\hypersetup{pdfborder = {0 0 0}}
\newcommand*{\todo}[1]{\textbf{\textcolor{red}{\textbf{TODO: #1}}}}

\begin{document}


\title{Clustering Semantic Predicates in the \\ Open Research Knowledge Graph}
%
%


\author{Omar Arab Oghli\orcidID{0000-0002-9092-9096} \and
Jennifer D'Souza\orcidID{0000-0002-6616-9509} \and
S\"oren Auer\orcidID{0000-0002-0698-2864}}
\authorrunning{Arab Oghli et al.}
%
\institute{TIB Leibniz Information Centre for Science and Technology, Hannover, Germany
\email{\{omar.araboghli,jennifer.dsouza,auer\}@tib.eu}}

\maketitle              
%


\begin{abstract}


When semantically describing knowledge graphs (KGs), users have to make a critical choice of a vocabulary (i.e. predicates and resources). The success of KG building is determined by the convergence of shared vocabularies so that meaning can be established. The typical lifecycle for a new KG construction can be defined as follows: nascent phases of graph construction experience terminology divergence, while later phases of graph construction experience terminology convergence and reuse. In this paper, we describe our approach tailoring two AI-based clustering algorithms for recommending predicates (in RDF statements) about resources in the Open Research Knowledge Graph (ORKG) \url{https://orkg.org/}. Such a service to recommend existing predicates to semantify new incoming data of scholarly publications is of paramount importance for fostering terminology convergence in the ORKG.

Our experiments show very promising results: a high precision with relatively high recall in linear runtime performance. Furthermore, this work offers novel insights into the predicate groups that automatically accrue loosely as generic semantification patterns for semantification of scholarly knowledge spanning 44 research fields.

\keywords{Content-based recommender systems \and Open research knowledge graph \and Artificial Intelligence \and Clustering algorithms.}
\end{abstract}

\section{Introduction}

Traditional, discourse-based scholarly communication in ``pseudo-digitized'' PDF format is being now increasingly transformed to a completely new representation leveraging semantified digital-born formats e.g. within the Open Research Knowledge Graph (ORKG)~\cite{auer2020improving} among other initiatives~\cite{aryani2018research,baas2020scopus,birkle2020web,dessi2020ai,fricke2018semantic,manghi_paolo_2019_3516918,wang2020microsoft}. This ``digital-first'' scholarly information representation is based on a fundamentally new information organization paradigm that creates and uses \textit{structured, fine-grained scholarly content}. Specifically, in the ORKG, scholarly communication is based on a large, interconnected knowledge graph (KG) of \textit{fine-grained scholarly content}. Such an information organization paradigm facilitates the evolution of scholarly communication from documents for humans to read towards human \textit{and} machine-readable knowledge with the aim of alleviating human reading cognitive tie-ups. To this end, the ORKG-based scholarly communication comprises a crucial machine-actionable unit of scholarly content in the form of \textit{human and machine-readable comparisons} of semantified \textit{scholarly contributions}~\cite{oelen2020generate}. These comparisons are meant to be used by researchers to quickly get familiar with existing work in a specific research domain. 
For example, determining the reproduction number estimate R0 of the Sars-Cov-2 virus from a number of studies in various regions across the world \url{https://orkg.org/comparison/R44930}. The semantically represented scholarly contribution comparisons in ORKG are especially necessary in our era of the deluge of peer-reviewed publications \cite{stm} and preprints~\cite{chiarelli2019accelerating} to help researchers stay on top of the fast-paced scientific progress. It concretely helps scientists to still keep an oversight over scientific progress by freeing unnecessary human cognitive tie-ups involved when searching for key information buried in large volumes of text.

The ORKG \textit{machine-readable comparisons} depend on the availability of a knowledge base of \textit{machine-actionable, semantified scholarly contributions}. The scholarly contributions are a unit of information defined in the context of the ORKG that describe the addressed problem and comprise the utilized materials, employed methods and yielded results in a scholarly article -- a model which subsumes \textsc{Leaderboard}s \cite{hou2019identification,kabongo2021automated}. A large community of researchers has recently been growing around the crowdsourced curation of \textit{scholarly contributions} in the ORKG (e.g., \url{https://orkg.org/paper/R163747}).\footnote{The related construct to ORKG contributions, of \textsc{Leaderboard}s in AI \url{https://paperswithcode.com/} has also garnered large-scale crowdsourcing interest.} To describe the scholarly contributions, RDF statements are used as structured semantic units that are machine-actionable as a result. A core semantic construct of these contribution-centric statements are the \textit{predicates} or \textit{properties} used to describe the contribution of an article. While the \textit{subject} and \textit{object} are content-based, \textit{predicates} can generically span contributions across articles. E.g., \textit{task name}, \textit{dataset name}, \textit{metric}, and \textit{score} are a group of four predicates used to semantically describe the leaderboard contribution across AI articles~\cite{kabongo2021automated} in the Computer Science domain; the predicates \textit{basic reproduction number}, \textit{confidence interval (95\%)}, \textit{location}, and \textit{time period} are used to describe Covid-19 reproductive number estimates in epidemiology articles~\cite{r0-covid}.

\textit{Predicates} are a core construct for semantically describing contributions in ORKG. To base the ORKG on meaningfully described semantic scholarly contributions, certain, specific groups of predicates that can capture key contribution aspects of the scholarly articles are essential. Each such group then becomes a \textit{contribution-centric} predicate group. Further, the group varies in applicability from being applicable to only a specific scholarly contribution or generalizing across a group of contributions from different papers. In this respect, the ORKG follows an agile, iterative Wiki-style collaboration approach giving curators the autonomy to coin new properties easily, but aims in the long-term trajectory to be coherent in terms of vocabulary for both predicates and resources. Note that contributions can only be compared based on standard predicates terminology for the \textit{machine-readable ORKG comparisons}. Further, the typical lifecycle of a new KG construction must also be accounted which starts with nascent phases of graph construction experiencing terminology divergence, while later phases of graph construction aim at terminology convergence and reuse. In this background setting of building the ORKG, the overarching research question investigated in this paper is: \textit{How to ensure that individuals, free to use arbitrary terminology, converge towards shared vocabularies for contribution-centric semantic predicates?}


Allowing users to make arbitrary statements is important, since it ensures that the expression of the diverse discoveries in Science are not being lost or unrepresented due to restricted semantic vocabularies. However, some authoring considerations need to be made. Without further considerations, the authoring freedom of contributions in the ORKG would result in statements with different vocabularies, defying the purpose of the need to semantify contributions. A terminology policy could be enforced but that would highly restrict users. Instead, a suggestion mechanism, recommending terminology based on the dataset, would help converge terminology without forcing users, as demonstrated in collaborative tagging \cite{lund2005social,marlow2006ht06}. In collaborative data entry, participants construct a dataset by continuously and independently adding further statements to existing data. Each curation participant faces the question: Which vocabulary elements to use? To ensure convergence, the answer is: use the most relevant and frequently occurring vocabulary elements. Finding the most frequent vocabulary elements is straightforward: one can simply count the occurrences. We therefore focus on finding the relevant vocabulary elements. Science comprises very heterogeneous contributions. Finding the vocabulary that is relevant for one contribution therefore means: finding similar contributions and reuse their vocabulary.

To this end, this work describes our implementation of an unsupervised AI service based on clustering similar papers and recommending contribution-centric predicate groups from the existing ORKG contributions. \textit{Similar scholarly contributions should be semantified with a homogeneous contribution-centric semantic predicate groups.} This is our intuition behind adopting clustering since the method aims to group the data points having similar features, where data points in different groups should have highly offbeat features. We chose hierarchical (Agglomerative\footnote{\url{https://scikit-learn.org/stable/modules/clustering.html\#hierarchical-clustering}}) and non-hierarchical (K-means\footnote{\url{https://scikit-learn.org/stable/modules/clustering.html\#k-means}}) clustering strategies. We avoid computationally intensive methods (e.g., Affinity) or methods that can handle only small cluster sizes (e.g., Spectral clustering). 

In summary, the contributions of our work are:
\begin{enumerate}
    \item a formalization of the application of homogeneous related groups of predicates to semantically describe scholarly contributions;
    \item the evaluation of two contrasting flavors of clustering objectives (hierarchical and non-hierarchical) to semantify contributions based on contribution-centric predicate groups. Since the task itself is formalized for the first time in this work, the application of an AI approach is correspondingly novel;
    \item detailed empirical evaluations of four machine learning model variants resulting from testing two different embedding representations; and
    \item the demonstration of a predicates recommendation service for the ORKG scholarly knowledge digitalization platform. Its objectives are two-fold: $i$) expedite adding a new contribution to the graph, and $ii$) semantify the contributions with a shared vocabulary. The recommender service takes as input a paper's title and abstract and in turn recommends a group of semantically related predicates based on earlier similar semantified papers if found by the clustering method, otherwise an empty set of predicates is returned. Such a system is described for the first time.
\end{enumerate}

The remainder of the paper is organized as follows. We first define the core concepts relevant to this work but which may be new in the community in Section 2. We then offer the formalized definition for our \textit{contributions-centric predicates grouping} task in Section 3, following which, in Section 4, we explain the custom dataset created from the ORKG RDF data dump incorporating our novel task. Next, we introduce our method for the contribution-centric predicates group recommendation service in Section 5. We then show the experimental results from our methods on our created custom task corpus in Section 6. Finally, we conclude with discussions on the possibility of further improvement and future work in Section 7.

\section{Definitions}
\label{sec:def}

We first define the central concepts to the task attempted in this work.


\paragraph{Contribution.} Highlights the findings of a research endeavour. An ORKG \textit{contribution} addresses a research problem, and can be described in terms of the materials and methods used and the results achieved. Contributions in different papers addressing the same research problem can be expected to have comparable semantic descriptions at least for their key properties whose values, i.e. resources and literals, then are specific to the research endeavour.

\paragraph{Contribution Triple.} Contributions are semantically described in a series of $(subject, predicate, object)$ RDF triple statements that build the ORKG. 

\paragraph{Contribution Predicates' Set (\textsc{cps}).} Is a set of predicates in contribution triples.

\paragraph{Comparison.} The ORKG supports downstream smart applications such as the creation of comparisons/surveys over its structured contributions. In other words, given the ORKG structured contributions, it is possible to compare the values of several such machine-actionable contributions provided their \textsc{cps}s are more or less similar. Comparisons can either be generated over several contributions of a single article (e.g., comparison of an AI benchmark characteristics having similar \textsc{cps}s but differing values over the different data domains annotated \url{https://orkg.org/comparison/R163843/}); or over contributions with similar \textsc{cps}s in different articles (e.g., comparison of the Covid-19 reproductive number (R0) estimate set of studies, respectively, conducted by different research groups for different countries \url{https://orkg.org/comparison/R44930/}).

\paragraph{Contribution Template.} An ORKG template is a set of predicates manually specified by a domain expert to describe contributions over a specific research problem. It helps standardize the process in ORKG of semantifying contributions with similar properties but different values. E.g., the leaderboard template \url{https://orkg.org/template/R107801}.


\paragraph{Contribution Predicates' Group (\textsc{cpg}).} A \textsc{cpg} is similar to a template. Where in templates, the predicates' groups are manually created by a domain expert, in the case of \textsc{cpg}s, they are heuristically obtained from several similar \textsc{cps}s. E.g., since in the process of obtaining the crowdsourced contributions in ORKG, it may surface that some \textsc{cps}s have been used similarly and repeatedly (say, above a designated threshold of contributions) to structure new contributions, they are then designated as a \textsc{cpg}. \textsc{cpg}s are indeed potential candidate templates. 

The goal of this work is to recommend \textsc{cpg}s discovered from \textsc{cps}s for describing new contributions, thereby offering ORKG users intelligent assistance in the process of structuring the new contributions. In the next section, we offer our actual task definition.

\section{Task Definition}
\label{sec:taskdef}

As mentioned above, the task addressed in this paper deals with discovering \textsc{cpg}s from \textsc{cps}s in the ORKG knowledge base (KB) to describe a new article's contribution. At a high-level, given an article and the ORKG KB of crowdsourced, structured contributions w.r.t. their \textsc{cps}s, the most relevant \textsc{cpg}, if found, should be recommended for describing the new article contribution. 

Our task formalism is as follows. The ORKG KB comprising structured contributions defined only w.r.t. predicates is $\textsc{CPS} = \{\textsc{cps}_1, ... ,\textsc{cps}_N\}$ which were used to structure contributions in the set $C = \{c_1, ... , c_N\}$, respectively. Here, the base $N$ represents the total number of contributions in the ORKG, \textsc{CPS} is the knowledge base of predicates sets, and $\textsc{cps}_i$ is the predicates set used to structure contribution $c_i$. Furthermore, the set of predicates in each \textsc{cps} formally is, $\textsc{cps}_i = \{p_{1i}, p_{2i}, ... , p_{xi}\}$. Finally, $P = \{p_1, p_2, ... , p_y\}$ represents the set of unique predicates aggregated from all \textsc{cps}s and $y$ is the total number of unique predicates used to structure the knowledge about contributions in ORKG. The recommendation task attempted in this work can then be defined as, given a new paper $P$ as its title $T$ and abstract $A$, to semantify or describe its contribution $C$ with an automatically discovered \textsc{cpg} from the ORKG KB of \textsc{cps}s such that the predicates in \textsc{cpg} $\in P$. 

\section{Task Dataset}
\label{sec:dataset}

For our novel task as defined above in \autoref{sec:taskdef}, a novel dataset needed to be created. Our raw data source was the ORKG RDF data dump \url{https://orkg.org/orkg/api/rdf/dump} dated 2021-11-10. Our objective with creating the dataset is to capture instances of the constructs of \textsc{cpg}s and \textsc{cps}s with their respective scholarly articles' title $T$ and abstract $A$. Instantiated \textsc{cpg}s in a dataset can serve two purposes: 1) a supervision signal for machine learning if building a supervised recommendation system; and 2) as gold-standard data to evaluate system automatic recommendations. While obtaining \textsc{cps}s and their corresponding articles may be a relatively straightforward process -- for \textsc{cps}s, we query the ORKG RDF data dump of contributions and for articles' $T$ and $A$, we query external services like Crossref -- the process of obtaining \textsc{cpg}s is not.

Recall in \autoref{sec:def}, our definition of ORKG \textit{Comparisons}. Briefly, they are a downstream application enabling computing surveys over collections of machine-actionable, structured contributions with roughly similar \textsc{cps}s. Given this and our need to obtain \textsc{cpg}s heuristically, we ask ourselves: could the \textsc{cps}s aggregated in \textit{Comparisons} be considered a \textsc{cpg}? The answer is ``yes,'' but with a caveat. We cannot consider the aggregated \textsc{cps}s as \textsc{cpg}s from just about any \textit{Comparison}. \textit{We want to generate \textsc{cpg}s that are strong candidates for templates}. For this, we deem that the candidate \textsc{cps}s need to demonstrate a strong repetition pattern of structuring several contributions as a determiner that they would apply to new contributions that have not yet been structured as well. Note here the connection between our heuristic and templates is that templates are defined with the intention of standardizing the process of structuring similar contributions on the same research problem across papers where they occur. Thus we concretely implement our heuristic as follows to generate \textsc{cpg}s for our dataset. The aggregated \textsc{cps}s in \textit{Comparison}s containing at least 10 contributions were considered as \textsc{cpg}s in our task dataset. 
To offer the reader better insights to the kind of \textit{Comparisons} that were finally considered, we show in \autoref{table:comparisons_cpgs} some comparisons from whose aggregated \textsc{cps}s, \textsc{cpg}s could be obtained. These \textit{Comparisons} include between 10 to at most 55 structured contributions. Since structured contributions in ORKG can span Science at large, the comparisons shown in \autoref{table:comparisons_cpgs} have a diverse coverage of research fields: 1 is from the Information Science, 2 belongs to Semantic Web, 3 belongs to Bioinformatics, 4 is from Urban Studies and Planning, 5 is from Software Engineering, and 6 belongs to Natural Language Processing.

\begin{table}[!ht]
\caption{Example ORKG Comparisons, which are aggregations of collections of structured contributions, and correspondingly some contribution predicates in their respective contribution predicates' groups (\textsc{cpg}).}
\label{table:comparisons_cpgs}
    \begin{tabular}{|l|p{5cm}|p{7cm}|}
    \hline
     & \textbf{Comparison title} & \textbf{Predicates in \textit{contribution predicate groups}}  \\
     \hline
     
    1. & Design and implementation of epidemiological surveillance systems \url{https://orkg.org/comparison/R146851} & epidemiological surveillance approach, epidemiological surveillance architecture, epidemiological surveillance software, epidemiological surveillance users, statistical analysis techniques \\
    \hline
    
    2. & Ontology learning from Folkosonomies \url{https://orkg.org/comparison/R144121} & learning method, and binary valued properties as: terms learning, concepts learning, individual learning, axioms learning \\
    \hline
    
    3. & Review of the existing research applying Deep Learning related to mental health conditions \url{https://orkg.org/comparison/R139050} & study cohort, used models, data, outcomes, outcome assessment method, performance \\
    \hline
    
    4. & Enterprise architecture applications for managing digital transformation of smart cities \url{https://orkg.org/comparison/R146458} & has methodology, issues addressed, study purpose, technology deployed \\
    \hline
    
    5. & Overview of Approaches that Classify User Feedback as Feature Request \url{https://orkg.org/comparison/R112387} & tf-idf precision, tf-idf recall, tf-idf f1, bag-of-words precision, bag-of-words recall, bag-of-words f1 \\ 
    \hline
    
    6. & NLP Datasets for Scientific Concept and Relation Extraction \url{https://orkg.org/comparison/R150058} & data domains, data coverage, dataset name, concept types, relation types, number of concepts, number of relations \\
    \hline
    
    \end{tabular}
\end{table}

Finally, our task dataset contains a set of structured contributions as \textsc{cps}s with their paper title $T$ and abstract $A$. Further, these structured contributions only pertain to those from which \textsc{cpg}s could be obtained heuristically from \textit{Comparisons} with the contribution included. The \textsc{cpg}s are also mappings to the original paper they roughly structure.

\begin{table}[ht!]
\centering
\caption{A tabular view of our task dataset statistics where the information in columns expressed w.r.t. the ORKG \textit{Comparisons} that were considered for the dataset.}
    \begin{tabular}{||c | c | c | c | c ||} 
     \hline
     - & Papers & Contributions & Predicates & Research Fields \\ [0.5ex] 
     \hline\hline
     Minimum per Comparison & 2 & 10 & 2 & 1 \\ [0.5ex] 
     \hline
     Maximum per Comparison & 202 & 250 & 112 & 5 \\ 
     \hline
     Average per Comparison & 23.25 & 35.47 & 12.86 & 1.19 \\ 
     \hline
     Total & 3941 & 5123 & 1681 & 44 \\ 
     \hline
    \end{tabular}
\label{table:cluster_data_statistics_by_comparisons}
\end{table}

\begin{figure}[ht!]
\centering
\begin{tabular}{cc}
    \includegraphics[width=0.5\textwidth]{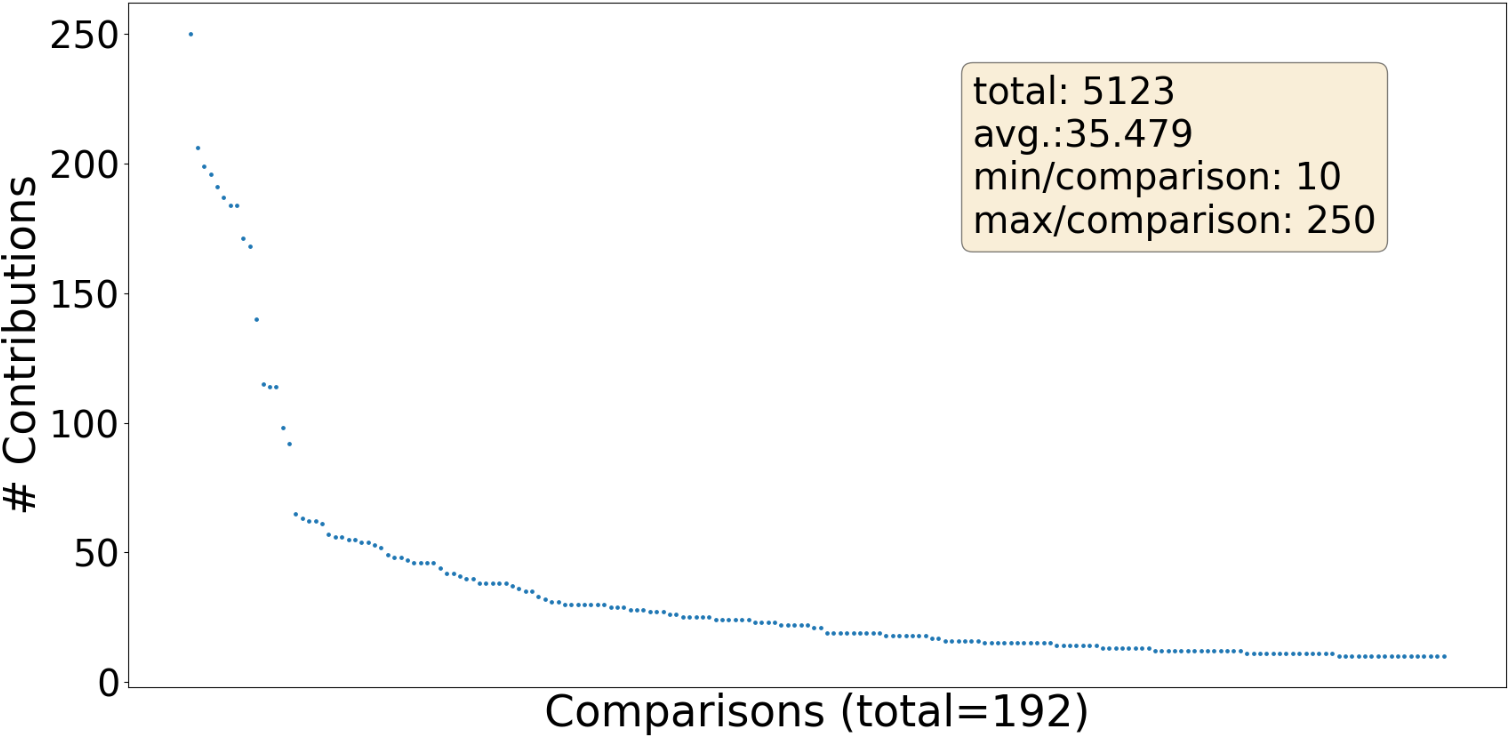} &
    \includegraphics[width=0.5\textwidth]{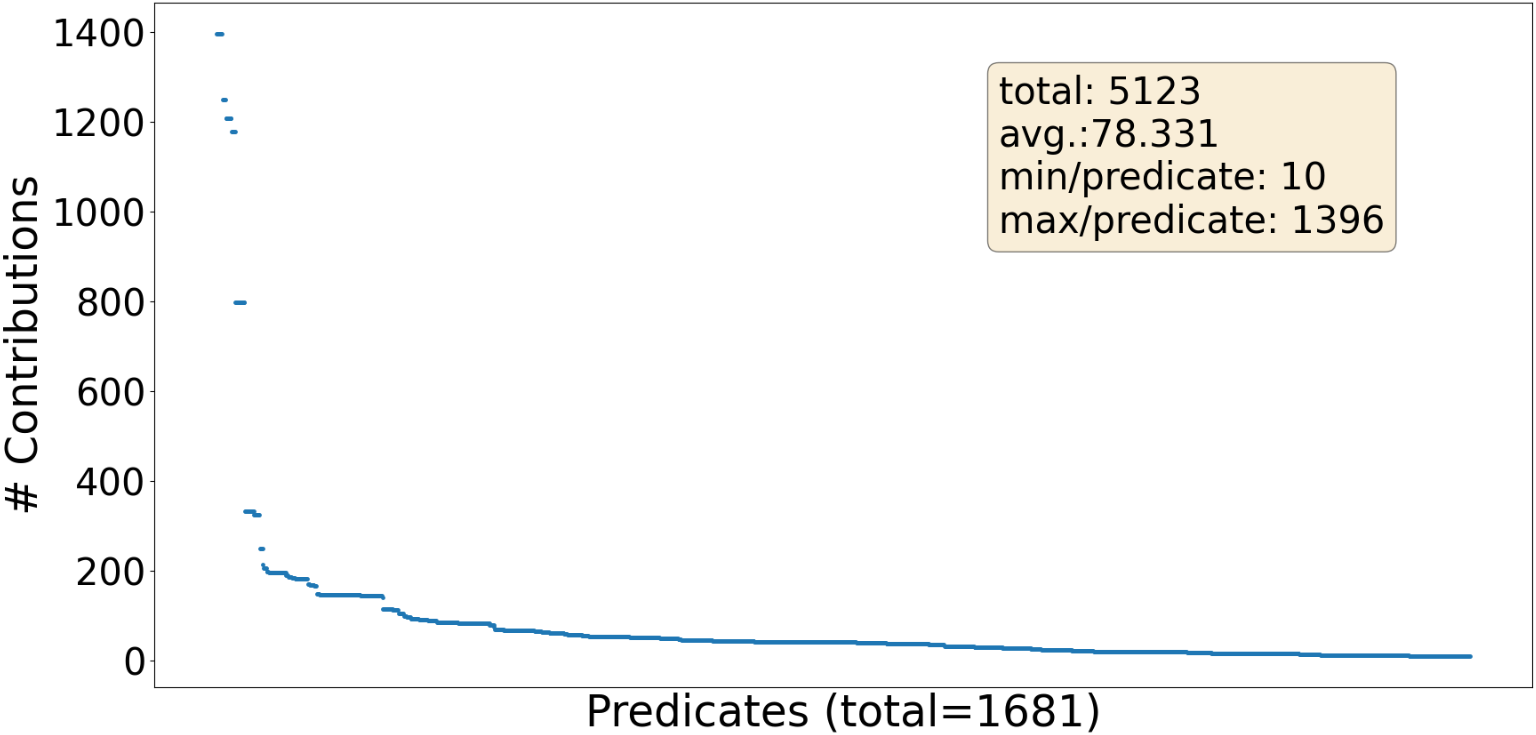}
\end{tabular}
\caption{Distribution of contributions in ORKG Comparisons in our task dataset (left) shown as a trendline. And the distribution of contributions over individual predicates in our task dataset shown as a trendline (right). This latter figure, in other words, shows the predicate repetition pattern for structuring contributions. }
\label{fig:clustered_data_papers_distrubtion}
\end{figure}

\paragraph{Dataset Statistics.} We now offer the reader some concrete statistical insights into our task dataset. \autoref{table:cluster_data_statistics_by_comparisons} shows the total unique papers, contributions, predicates and their research fields' coverage. The minimum, maximum, and average numbers are aggregated by the selected \textit{Comparisons} from which \textsc{cpg}s could be obtained. Thus our task dataset includes 3941 papers and 1681 unique predicates. The selected \textit{Comparisons} have included 23.25 papers on average, with a minimum of 2 and maximum of 202. Further, contributions were structured by as few as 2 predicates and as many as 112 predicates at an average rate of 12.86. Our corpus covers contributions from across 44 different research fields. \autoref{fig:clustered_data_papers_distrubtion} depicts the trendline patterns of the distribution of contributions in ORKG \textit{Comparisons} and the distribution of predicates to structure the contributions. We see they are respectively a long-tailed distribution, i.e. some comparisons are outliers in our dataset and include a large number of contributions, and on the other hand, some predicates are used most frequently to structure nearly most of the contributions. Our task dataset is publicly available at \url{https://doi.org/10.5281/zenodo.6513499}. 

\section{Background and Related Work}
The Semantic Web~\cite{1637364} was first introduced as an extension to the World Wide Web as a comprehensive framework to create meaningful (semantic) description of information on the web as subject-predicate-object triples logical equivalent of unstructured text. Different Semantic Web languages (such as RDF, RDFS and OWL) were introduced and are now widely used standards to create structured data for intelligent computational agents~\cite{Fernandez2018}. 

Over the past decade, scholarly knowledge has started gaining traction for representation in machine-interpretable form leveraging Semantic Web standards. In other words, to semantically describe research knowledge gained among researchers from around the world. To this end, various initiatives are focused on constructing \textit{Knowledge Graphs} (KGs) over different aspects of scholarly knowledge. One line of work concerns metadata-interlinking (e.g. authors, citations or keywords)~\cite{10.1007/978-3-030-30796-7_8,yaman_et_al:OASIcs:2019:10379,ammar-etal-2018-construction,lo-etal-2020-s2orc}. Another line of work~\cite{Aryani2017,BECHHOFER2013599,manghi_paolo_2019_2643199} supports the construction of KGs based on interlinking of research artifacts (e.g. source code, datasets or figures). In contrast, the ORKG~\cite{auer2020improving} aims for contribution-centric interlinking of research resources.

Constructing a KG remains a challenge, in general, w.r.t. ensuring the graph quality and graph knowledge completeness. Four main groups of construction methods were classified by~\cite{7358050} as $i)$ expert-based manual curation~\cite{10.1145/219717.219745,10.1145/219717.219748,Bodenreider2004-vx}, $ii)$ open community-based manual collaborative curation~\cite{10.1145/1376616.1376746,10.1145/2629489}, $iii)$ automated semi-structured approaches involving information extraction from structured tabulated data or using rule-based systems~\cite{10.1007/978-3-540-76298-0_52,10.1145/1376616.1376746}, and $iv)$ automated unstructured approaches using machine learning to mine text~\cite{45634,10.5555/2390948.2391076,10.1145/1935826.1935869,10.4018/jswis.2012070103,10.5555/2898607.2898816}. In fact, the ORKG reflects all four of the information curation approaches. For instance, the creation of ORKG templates by domain experts as reusable graph patterns is an expert-based manual curation approach; the crowdsourcing of research contributions in the ORKG is a community-based collaborative approach~\cite{lund2005social,marlow2006ht06}. As automated semi-structured methods, often rule-based approaches are also experimented with in the context of the ORKG, such as the system to acquire scientific entities from scholarly article titles \cite{10.1007/978-3-030-91669-5_31}. Finally, the ORKG's fully automated text mining include the Leaderboard extraction system~\cite{10.1007/978-3-030-91669-5_35} and Computer Science Named Entity Recognition~\cite{https://doi.org/10.48550/arxiv.2203.14579}.


The predicates recommendation system described in this paper would then be a service offered for community-based curation of the ORKG to ensure repetition of the contribution graph patterns toward terminology convergence and comparability of the structured descriptions. Relatedly then, in terms of approaches, \cite{10.1007/978-3-540-72667-8_13} introduced two simple algorithms. One based on generic similarity metrics~\cite{BRODER19971157,web_metrics}, containment and resemblance, used to classify predicates similar to the query resource as a ranked list. And another approach based on computing co-occurrence matrix of predicates of resources. Despite the linear runtime performance of the algorithms, they rely on a structured query resource, while we are in need of a predicate suggestion service based on unstructured texts. To this end, \cite{bioassays} offered a clustering methodology using K-means~\cite{Jin2017} to semantify descriptions of Biological Assays. Our approach is modelled after this system. Finally, while at first glance, topic modeling~\cite{blei2003latent} may seem similar to a clustering method, we observe that topic modeling would help us obtain distribution of a paper over topics, whereas our objective is instead to assign a paper to a single semantic group informed by its contribution, which we achieve via clustering where each paper is assigned to only a single cluster. Nevertheless, to ensure experimental completeness, we show results from a topic modeling baseline.

\section{Our Approach}

\subsection{Clustering of Contribution Predicates' Groups}

Earlier in the task dataset section (\autoref{sec:dataset}), we first described our heuristic reliance on ORKG \textit{Comparisons} to obtain \textsc{cpg}s. The next question is: how can we develop a recommender of \textsc{cpg}s given a corpus of papers as their titles $T$ and abstracts $A$ structured for their contributions with \textsc{cps}s? We propose an AI-based unsupervised clustering strategy of papers as the solution. With this approach, we aim to automatically obtain \textsc{cpg}s by aggregating all \textsc{cps}s in a particular cluster of similar papers. Our hypothesis is that papers describing similar contributions are also similar to each other in terms of their unstructured text descriptions as $T$ and $A$. Thus, the role played by the construct of \textit{Comparison}s in generating \textsc{cpg}s are now replaced, in the context of an automated recommender, by a clustering algorithm.


\subsection{Grouped Predicates Recommender System Workflow}

Our automated recommender system workflow with clustering is as follows.


\begin{enumerate}
    \item A user provides the paper's title and/or DOI they wish to add.
    \item We fetch the paper's abstract from external service APIs as discussed in the dataset section (\autoref{sec:dataset}).
    \item The paper's title and abstract are concatenated and vectorized.
    \item The vector representation will be fed to a pre-trained clustering model for most relevant cluster prediction. Note, each candidate cluster was constructed based on prior semantified papers already in the ORKG KB.
    \item We fetch the predicates, i.e. \textsc{cps}s, of all the structured paper contributions in the predicted cluster.
    \item All \textsc{cps}s are combined to a set to produce a \textsc{cpg} which is then recommded to the user for their query paper.
\end{enumerate}

This workflow is illustrated in \autoref{fig:predicates_recommendation_service_workflow} below.

\begin{figure}[ht!]
    \centering
    \frame{\includegraphics[width=\textwidth]{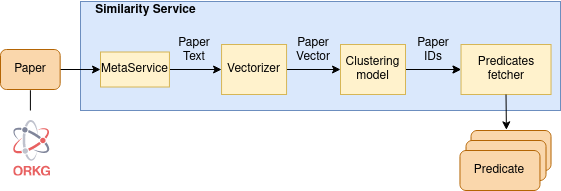}}
    \caption{Our grouped predicates recommender workflow. Arrows indicate the data flow.}
    \label{fig:predicates_recommendation_service_workflow}
\end{figure}

For the vectorizer module, we experimented with two different vectorization functions. And, for the clustering model, we tried two different clustering algorithms. These experimental details are provided next.

\subsection{Vectorization Functions}

We rely on two vectorization methods. 1) TF-IDF - vectors are directly computed from our task corpus. 2) SciBERT~\cite{beltagy2019scibert} - vectors are computed from a pretrained model of embeddings over large-scale publications' data.

\paragraph{\textsc{TF-IDF} embeddings -} We use the scikit-learn~\cite{sklearn_api,scikit-learn} library to convert our corpus of paper $T$ and $A$ into TF-IDF~\cite{tf_idf_embeddings} vectors. TF-IDF vectors are $n$-dimensional real-valued vectors representing a given text with the term frequency-inverse document frequency (TF-IDF) value for each possible term in the corpus. $260,016$ unique terms were found in our corpus.

\paragraph{\textsc{SciBERT} embeddings -} We feed forward the pre-trained AllenNLP SciBERT~\cite{beltagy2019scibert} uncased model with our text corpus of paper $T$ and $A$ to output its final hidden state, which is then averaged via sentence transformers (\url{https://huggingface.co/sentence-transformers}) resulting in a vectorized text of dimension $768$. We obtain the embeddings using a max sequence length of 512.

\subsection{Clustering Algorithms}

We rely on two complementary variants of clustering methods: one based on non-hierarchical clustering, specifically K-means; and another based on a hierarchical clustering, specifically agglomerative clustering. 

\paragraph{\textsc{K-means}.} Following~\cite{bioassays}, we apply the centroid-based clustering algorithm K-means~\cite{Jin2017} to group similar scholarly contributions represented by their paper $T$ and $A$. The scikit-learn implementation (\url{https://scikit-learn.org/stable/modules/generated/sklearn.cluster.KMeans.html}) was leveraged and the models were trained on the \textit{Google Colab Pro+} platform due to the complex time and space requirements of K-means.

\paragraph{\textsc{Agglomerative}.} The agglomerative bottom-up hierarchical-based clustering algorithm~\cite{Zepeda-Mendoza2013} with \textit{ward} linkage was applied. This method, like the K-means objective function, minimizes the variance within a cluster. Again, the scikit-learn (\url{https://scikit-learn.org/stable/modules/generated/sklearn.cluster.AgglomerativeClustering.html}) implementation was used and the models were trained on \textit{Google Colab}.

\subsection{Experimental Setup}
\label{exp-setup}

In this section, we describe our experimental setup to find the optimal vectorization method and clustering algorithm combination. 

\paragraph{Dataset.} First, we created training and test dataset splits of our task corpus. From each comparison, we split its papers in the 70:30 ratio for creating training and test datasets, respectively. The test dataset was reserved as a blind set with which the trained algorithm was queried for its predictions of clustered predicate groups. In total, our training set consisted of 3,696 contributions distributed over 192 comparisons, whereas our test set had 1,427 contributions distributed over 167 comparisons. The training and test sets contain mutually unique instances.

\paragraph{Evaluation Metrics.} Per the standard evaluation practice of information retrieval systems, we employed the macro- as well as the micro-average~\cite{Asch2013MacroandME} of the precision ($P$), recall ($R$) and F-score ($F1$).

\paragraph{Selecting K.} $K$ was strategically chose in the range  $|C| \leq k \leq |P|$ with a step size of 50, where $C=200$ is the set of ORKG comparisons and $P=2050$ is the set of training papers. 38 total models were obtained per vectorization method.

\paragraph{Predictions.} Some considerations need to be taken w.r.t. evaluating our clustering models. We put emphasis on the absence of the prediction function in the agglomerative algorithm compared to its presence in K-means that can simply assign a new incoming data instance to one of the clusters based on the distance to the centroid. In hierarchical clustering on the other hand, assigning a new data instance can entirely change the clusters because it can trigger several mergings based on the linkage measure. In order to avoid re-building the hierarchical clusters for each test instance, we build them only once on the entire dataset and evaluate by comparing the comparisons' predicates of the training papers included in the cluster to which a test instance is assigned with the expected ones. 

\section{Results and Discussion}

In this section, we discuss our experimental results for selecting the optimal vectorization and clustering model pair.

\subsection{Quantitative Evaluations}


\subsubsection{Baselines.} We implemented two baselines each driven by a research question (RQ). \textbf{\textit{Baseline 1 RQ}}: what happens if the problem were reduced to a trivial solution where clusters of contributions are created simply based on the research field? For this baseline, the contribution \textsc{cps}s in our training data were grouped to form a \textsc{cpg} per research field of the training data contributions. The 44 different research fields in our dataset thus resulted in 44 \textsc{cpg}s. Thus a new incoming paper from the test dataset would be assigned the \textsc{cpg} of its research field created from the training dataset contributions. Row 1 in \autoref{table:overall_results} shows the results from this baseline. We find that while a perfect recall can be obtained, such an approach is not precise. This reveals an important characteristic of our dataset: i.e., \textit{the structure of contributions within each research field can differ significantly across papers in the same field}. \textbf{\textit{Baseline 2 RQ}}: what happens when 192 topics are generated from our dataset by topic modeling~\cite{blei2003latent} analogous to the 192 comparisons? To implement this baseline, topic distributions were obtained for all papers in the training dataset and each paper was assigned to the best topic. Thus \textsc{cps}s were obtained per topic from which \textsc{cpg}s were generated. A new incoming test paper was then classified to best topic and assigned its \textsc{cpg}. Row 2 in \autoref{table:overall_results} shows the results from a topic modeling based approach. The results prove to not be promising in terms of both precision and recall. This is contrary to our initial assumption that topic groups could be a correlated semantic construct of comparisons. We find no correlation can be established.

\begin{table}[ht!]
\footnotesize
\centering
\caption{Results of automatically generating contribution predicates groups using K-Means clustering. $K$ was chosen in the range from 200 to 2050 in step sizes of 50. The table shows the most significant results obtained in terms of $P$, $R$, and $F1$.}
    \resizebox{\textwidth}{!}{\begin{tabular}{||c | c | c | c || c | c | c || c | c | c || c | c | c||} 
     \hline 
     - & \multicolumn{6}{c||}{Macro-Average} & \multicolumn{6}{c|}{Micro-Average} \\ [0.5ex] 
     \hline
     - & \multicolumn{3}{c||}{TF-IDF} & \multicolumn{3}{c||}{SciBERT} & \multicolumn{3}{c||}{TF-IDF} & \multicolumn{3}{c||}{SciBERT} \\ 
     \hline
     \thead{Clusters\\$K$} & $P$ & $R$ & $F1$ & $P$ & $R$ & $F1$ & $P$ & $R$ & $F1$ & $P$ & $R$ & $F1$ \\ 
     \hline\hline
        \hline
        200 & 0.344 & \underline{\textbf{0.957}} & 0.506 & 0.380 & 0.921 & 0.538 & 0.057 & \underline{\textbf{0.953}} & 0.108 & 0.242 & 0.913 & 0.383 \\
        \hline
        350 & 0.453 & 0.917 & 0.607 & 0.466 & \underline{\textbf{0.924}} & 0.620 & 0.272 & 0.906 & 0.419 & 0.320 & \underline{\textbf{0.919}} & 0.475 \\
        \hline
        1100 & 0.632 & 0.868 & \underline{\textbf{0.731}} & 0.625 & 0.881 & 0.731 & 0.447 & 0.838 & 0.583 & 0.480 & 0.886 & 0.623 \\
        \hline
        1650 & \underline{\textbf{0.650}} & 0.821 & 0.726 & 0.681 & 0.855 & 0.758 & 0.535 & 0.799 & 0.641 & 0.588 & 0.832 & 0.689 \\
        \hline
        1850 & 0.649 & 0.779 & 0.708 & 0.704 & 0.849 & 0.770 & \underline{\textbf{0.593}} & 0.732 & \underline{\textbf{0.655}} & 0.603 & 0.834 & 0.700 \\
        \hline
        2050 & 0.609 & 0.748 & 0.672 & \underline{\textbf{0.728}} & 0.844 & \underline{\textbf{0.781}} & 0.486 & 0.696 & 0.572 & \underline{\textbf{0.659}} & 0.808 & \underline{\textbf{0.726}} \\
        \hline
    \end{tabular}}
\label{table:clustering_evaluation_results}
\end{table}

\begin{table}[ht!]
\footnotesize
\centering
\caption{Results of automatically generating contributions predicates groups using Agglomerative clustering. $K$ was chosen in the range from 200 to 2050 in step sizes of 50. The table shows the most significant results obtained in terms of $P$, $R$, and $F1$.}
    \resizebox{\textwidth}{!}{\begin{tabular}{||c | c | c | c || c | c | c || c | c | c || c | c | c||} 
     \hline 
     - & \multicolumn{6}{c||}{Macro-Average} & \multicolumn{6}{c|}{Micro-Average} \\ [0.5ex] 
     \hline
     - & \multicolumn{3}{c||}{TF-IDF} & \multicolumn{3}{c||}{SciBERT} & \multicolumn{3}{c||}{TF-IDF} & \multicolumn{3}{c||}{SciBERT} \\ 
     \hline
     \thead{Clusters\\$K$} & $P$ & $R$ & $F1$ & $P$ & $R$ & $F1$ & $P$ & $R$ & $F1$ & $P$ & $R$ & $F1$ \\ 
     \hline\hline
        \hline
        200 & 0.521 & \underline{\textbf{0.970}} & 0.678 & 0.309 & 0.031 & 0.056 & 0.160 & \underline{\textbf{0.979}} & 0.275 & 0.198 & 0.032 & 0.055 \\
        \hline
        250 & 0.550 & 0.967 & 0.701 & 0.354 & \underline{\textbf{0.031}} & \underline{\textbf{0.057}} & 0.189 & 0.977 & 0.317 & 0.265 & 0.032 & 0.057 \\
        \hline
        350 & 0.592 & 0.955 & 0.731 & 0.390 & 0.030 & 0.056 & 0.239 & 0.964 & 0.383 & 0.312 & \underline{\textbf{0.032}} & \underline{\textbf{0.058}} \\
        \hline
        1100 & 0.811 & 0.869 & \underline{\textbf{0.839}} & 0.621 & 0.022 & 0.042 & 0.736 & 0.875 & 0.799 & 0.648 & 0.023 & 0.045 \\
        \hline
        1300 & 0.823 & 0.845 & 0.834 & 0.751 & 0.021 & 0.041 & 0.760 & 0.853 & \underline{\textbf{0.804}} & 0.761 & 0.023 & 0.044 \\
        \hline
        1950 & 0.869 & 0.743 & 0.801 & \underline{\textbf{0.823}} & 0.013 & 0.026 & 0.830 & 0.735 & 0.779 & \underline{\textbf{0.828}} & 0.012 & 0.024 \\
        \hline
        2000 & \underline{\textbf{0.874}} & 0.733 & 0.797 & 0.823 & 0.013 & 0.026 & \underline{\textbf{0.835}} & 0.727 & 0.777 & 0.828 & 0.012 & 0.024 \\
        \hline
    \end{tabular}}
\label{table:clustering_evaluation_results_agglomerative}
\end{table}

\begin{table}[ht!]
\centering
\caption{Overall results - Comparison between $Base_{RF}$ (Baseline Research Fields), $Base_{LDA}$ (Baseline Latent Dirichlet Allocation), K-Means and Agglomerative.}
    \begin{tabular}{||c | c | c | c || c | c | c || c | c | c || c | c | c||}
     \hline 
     - & \multicolumn{3}{c||}{Macro-Average} & \multicolumn{3}{c|}{Micro-Average} \\ [0.5ex] 
     \hline
     \thead{Approach} & $P$ & $R$ & $F1$ & $P$ & $R$ & $F1$ \\ 
     \hline\hline
    $Base_{RF}$ & 0.186 & 1.0 & 0.250 & 0.028 & 1.0 & 0.055 \\
    \hline
    $Base_{LDA}$ & 0.040 & 0.662 & 0.090 & 0.023 & 0.615 & 0.046 \\
    \hline
    K-Means & 0.728 & 0.844 & 0.781 & 0.659 & 0.808 & 0.726 \\
    \hline
    Agglomerative & 0.823 & 0.845 & 0.834 & 0.760 & 0.853 & 0.804 \\
    \hline
    \end{tabular}
\label{table:overall_results}
\end{table}

\subsubsection{Clustering Results.} 

Tables \ref{table:clustering_evaluation_results} and \ref{table:clustering_evaluation_results_agglomerative} show the results from applying K-means and Agglomerative clustering, respectively, with both tables showing results of the two vectorization methods. The best results are highlighted as bold and underlined in the respective tables. 

The evaluation results point out that each clustering method prefers a different vectorization strategy. The K-means clustering algorithm (see \autoref{table:clustering_evaluation_results}) show that SciBERT embeddings are the preferred vectorization method obtaining $0.726$ micro $F1$ and $0.781$ macro $F1$ (k = 2050). The Agglomerative clustering algorithm (see \autoref{table:clustering_evaluation_results_agglomerative}) show that TF-IDF embeddings is the preferred vectorization method obtaining $0.804$ micro $F1$ and $0.834$ macro $F1$ (k = 1300). Thus, Agglomerative clustering surpasses K-means by nearly 10 points. While macro scores are evaluations at the Comparisons level, micro scores report evaluations at a more fine-grained predicates level. Based on this, our optimal model is at $k=1300$ with the highest micro $F1$ using TF-IDF vectorization and Agglomerative clustering.

\subsection{Qualitative Evaluations}

We qualitatively analyze the ability of our system to regenerate ORKG \textit{Comparisons} via clustering. In other words, are the contributions in ORKG \textit{Comparisons}, even if contained in different clusters, distributed over pure or impure clusters? A pure cluster is one that contains contributions from only a single \textit{Comparison}; an impure cluster is one that contains contributions from multiple \textit{Comparisons}. We define this measure for regenerating the ORKG comparisons automatically as follows.


\begin{equation}
    ReGen(comp) = \frac{|\{c \in C \ | \ c \ groups \ papers \ only \ from \ comp\}|}{|C|}
    \label{equation:purity}
\end{equation}


We pick a representative example in our qualitative analysis. The ReGen value for ``Smart cities and cultural heritage'' ORKG \textit{Comparison} which has 4 contributions originally (\url{https://www.orkg.org/orkg/comparison/R140131/}) is 50\%. As a result of our method, this Comparison spanned 4 clusters (2 pure and 2 impure). However, observing the impure clusters closely, we noted that they included contributions from other \textit{Comparisons} (e.g., ``Smart city governance research categories analysis by references articles'' or ``Enterprise architecture applications for managing digital transformation of smart cities'', etc.) which were on the same research theme of ``Smart Cities'' and therefore had more or less very similar predicates. Thus, having not perfectly regenerable \textit{Comparisons} by our method does not necessarily imply inaccurate predicted clusters. But this finding points to the fact that ORKG \textit{Comparisons} are not all necessarily too semantically distinct from each other. 

\section{Conclusion and Future Work}
Our experiments on the hierarchical Agglomerative algorithm have shown a quantitative result of 80.4\% F1 and a qualitative result of similar recommendations of comparison predicates to those predefined in ORKG templates. Thus, the content-based recommender system based on clustered predicate units satisfies the templating concept of the ORKG. Overall, we offer among our methodology a semantification system for research contributions in the Semantic Web that does not limit the user autonomy, but instead directs the user to choose from an existing vocabulary, and hence prevent terminology divergence during later phases of graph construction.

As future work, the method will continue to be retrained for its clusters based on the ever-growing ORKG KB. Also, it is planned to implement a better association that is not heuristic-based to determine if indeed clustering related predicates produces templates.


\paragraph*{Supplemental Material Statement:}
Dataset leveraged for constructing the clusters is available from \url{https://doi.org/10.5281/zenodo.6513499}. The code base for both training and evaluation is available from \url{https://doi.org/10.5281/zenodo.6514139}. Please check the publication descriptions for further details.

%
%
%
%
\bibliographystyle{template/splncs04}
\bibliography{mybibliography}

\end{document}